\documentclass[letterpaper, 10 pt, journal, twoside]{IEEEtran}  % Comment this line out if you need a4paper

\IEEEoverridecommandlockouts                              % This command is only needed if 
\usepackage{amsthm,amsmath,amssymb}
\usepackage{mathrsfs}
\usepackage{graphicx}
\usepackage{float}
\usepackage{cite}
\usepackage{multirow}
\usepackage{diagbox}
\usepackage{stfloats}
\usepackage{multicol}
\usepackage[noend]{algpseudocode}
\usepackage{algorithmicx}
\usepackage{algorithm}
\usepackage{inputenc}
\usepackage{subfigure}
\usepackage{bm}

\usepackage[switch]{lineno}  % 用于增加行号

\algnewcommand\algorithmicswitch{\textbf{switch}}
\algnewcommand\algorithmiccase{\textbf{case}}
\algnewcommand\algorithmicassert{\texttt{assert}}
\algnewcommand\Assert[1]{\State \algorithmicassert(#1)}%
\algdef{SE}[SWITCH]{Switch}{EndSwitch}[1]{\algorithmicswitch\ #1\ \algorithmicdo}{\algorithmicend\ \algorithmicswitch}%
\algdef{SE}[CASE]{Case}{EndCase}[1]{\algorithmiccase\ #1}{\algorithmicend\ \algorithmiccase}%
\algtext*{EndSwitch}%
\algtext*{EndCase}%

\title{Collaborative Target Tracking in Elliptic Coordinates: a Binocular Coordination Approach
}

\author{
	Yuan Chang$^{1}$, Zhiyong Sun$^{2}$, Han Zhou$^{1}$, Xiangke Wang$^{1}$, Lincheng Shen$^{1}$, Tianjiang Hu$^{3*}$% <-this % stops a space

\thanks{$^{*}$ Corresponding author (e-mail: hutj3@mail.sysu.edu.cn).}%
\thanks{$^{1}$ National University of Defense Technology, Changsha 410073, China.}%
\thanks{$^{2}$ Eindhoven University of Technology, Den Dolech 2, 5612 AZ, Eindhoven, the Netherlands.}%
\thanks{$^{3}$ Sun Yat-sen University, Guangzhou 510725, China.}%
}

\begin{document}

% \linenumbers   % 用于增加行号

\maketitle
% \thispagestyle{empty}
% \pagestyle{empty}

%%%%%%%%%%%%%%%%%%%%%%%%%%%%%%%%%%%%%%%%%%%%%%%%%%%%%%%%%%%%%%%%%%%%%%%%%%%%%%%%
\begin{abstract}

This paper concentrates on the collaborative target tracking control of a pair of tracking vehicles with formation constraints. The proposed controller requires only distance measurements between tracking vehicles and the target. Its novelty lies in two aspects: 1) the elliptic coordinates are used to represent an arbitrary tracking formation without singularity, which can be deduced from inter-agent distances, and 2) the regulation of the tracking vehicle system obeys a binocular coordination principle, which simplifies the design of the control law by leveraging rich physical meanings of elliptic coordinates. The tracking system with the proposed controller is proven to be exponentially convergent when the target is stationary. When the target drifts with a small velocity, the desired tracking formation is achieved within a small margin proportional to the magnitude of the target's drift velocity. Simulation examples are provided to demonstrate the tracking performance of the proposed controller.

\end{abstract}

% Keywords appear just beneath the abstract. Use only for final RAL version. 
%\begin{IEEEkeywords}
%	Multi-Robot Systems, Formation Control, Collaborate Target Tracking.
%\end{IEEEkeywords}
%%%%%%%%%%%%%%%%%%%%%%%%%%%%%%%%%%%%%%%%%%%%%%%%%%%%%%%%%%%%%%%%%%%%%%%%%%%%%%%%
\section{Introduction}

This paper investigates the collaborative target tracking problem and proposes a coordinated control method for a pair of tracking vehicles so that the tracking vehicles and the target eventually form a desired formation. The available measurements are inter-vehicle distances. Both the target representation and the design of the control law are established in the elliptic coordinate system, which encapsulates the coupled properties between the interacting vehicles, thereby simplifying the controller form. 

Cooperative target tracking (or so-called circumnavigation/encirclement control) has been studied extensively in the literature \cite{Daniel,sun_collaborative,shames_circumnavigation,Yu_circumnavigation}. The paper \cite{Daniel} proposed a general framework that consists of an outer loop and an inner loop. The outer loop is to design a reference velocity to match the movement of the target, and the inner loop is to control each agent so that the formation centroid can achieve the desired reference velocity. Recently, the paper [2] discussed collaborative target tracking using fixed-wing UAV systems with constant-speeds, in tracking a moving target with constant or time-varying speeds. Their works are based on the assumption that the target position and velocity are known, and the tracking controller demands relative position and heading measurements. In comparison, \cite{Yu_circumnavigation} designed a bearing-based circumnavigation approach, where the relative distance and velocity of the target is obtained by estimation. In this paper, we propose a binocular-based tracking control approach, which does not require the knowledge of estimation of target' positions, and thus greatly relaxes the assumptions on inter-vehicle measurements from the above mentioned works.

There are two issues that need to be addressed in cooperative target tracking with a binocular coordination approach: the first is how to represent formation constraints of the system constructed by binocular vehicles and the target, while the second is how to design a control law for the vehicles to eliminate the tracking errors. As such, several works on formation control have been addressed based on different available measurements \cite{distance_formation,angle_formation,mixed_angle_distance_formation}. One of the most popular methods is distance-based formation control \cite{distance_formation}, where the formation is achieved by adjusting inter-agent distances. 
In contrast, Bishop et al. proposed an angle-based formation control \cite{angle_formation}, where the control law is designed with subtended angles. Later, they extended the angle-based method to mixed-angle-distance formation \cite{mixed_angle_distance_formation} that involves scaling control. Our work differs from \cite{distance_formation,angle_formation,mixed_angle_distance_formation} in that the target does not actively participate in the formation control. Moreover, the proposed controller obeys a binocular coordination principle \cite{chang}, which treats binocular nodes as a joint system to design its tracking trajectory.

The contributions of this paper are twofold. First, we propose to describe an arbitrary tracking formation without singularity in elliptic coordinates, which can be viewed as a natural extension of polar coordinates for a binocular system. Second, we present a tracking control law with a succinct form by making full use of physical meanings and geometries of elliptic coordinates. We have proved that, under the proposed control law, the tracking system is exponentially convergent when the target is stationary. Moreover, it is robust to track a slow-drifting target in the sense that the tracking errors converge to a small margin proportional to the drift velocity. 

\section{Preliminary and problem statement}
\subsection{Binocular target tracking}

We consider the task of tracking a target with a binocular system. The considered binocular system consists of 2 tracking vehicles whose dynamics are described by a single integrator: 
\begin{equation}
\dot p_i =u_i,i \in \left\{ 1,2 \right\},
\label{eq_agentmodel}
\end{equation}
where ${{p}_{i}}\in {{\mathbb{R}}^{2}}$ and ${{u}_{i}}\in {{\mathbb{R}}^{2}}$ represent the position and control input of vehicle $i$, respectively. First, the following assumption is naturally established.

\noindent \textbf {Assumption 1:} The initial positions of the two tracking vehicles are non-coincident, i.e. $p_1(t_0) \neq p_2(t_0)$.

The measurements from vehicle $j$ available to vehicle $i$ are the relative distance measurement $d_{ij}=\left\| {{p}_{j}}-{{p}_{i}} \right\|$ and relative bearing measurement $g_{ij}=(p_j-p_i)/d_{ij}$. The target position $p_t$ is not directly accessible. Each vehicle $i$ only measures the relative distance $d_{it}=\left\| {{p}_{t}}-{{p}_{i}} \right\|$ to the target. Besides, the tracking vehicles only know which side of the baseline the target is located on, but do not have access to accurate bearing information.

\noindent \textbf {Remark 1.} We do not impose any restrictions on the target location. This is different from that of \cite{angle_formation}, which assumes that the tracking vehicles and the target cannot be collinear.

So far, the triangular formation in the tracking task can be uniquely represented by $d_{ij}$ and $d_{it}$. In the following subsection, we further electorate formation constraints with transformations of elliptic coordinates to facilitate the design of tracking control law.

\begin{figure}
	\centering
	\includegraphics[width=3.3in]{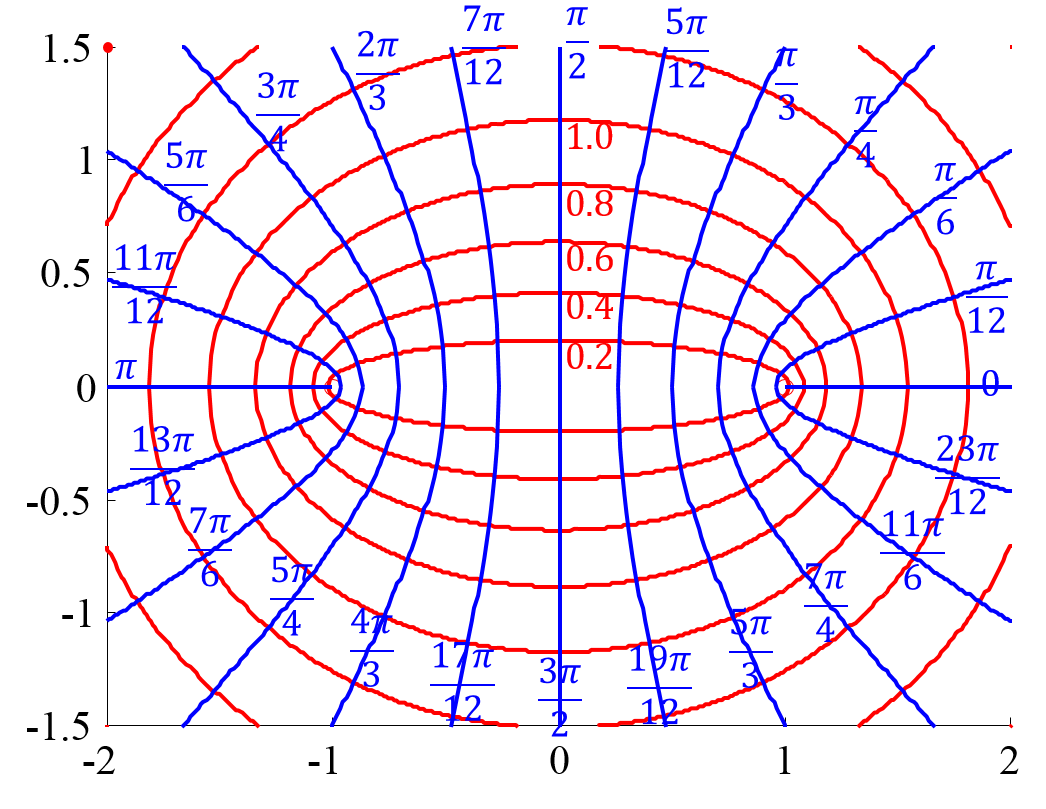}
	\caption{Depiction of an elliptic coordinate system with $c=1$. The red curves indicate ellipses with a constant $\xi$, while the blue curves indicate hyperbolae with a constant $\eta$.}
	\label{geometry}
\end{figure}

\subsection{Formation constraints in elliptic coordinates}

The elliptical coordinate system is a two-dimensional orthogonal coordinate system that has been commonly used for modeling and solving problems with binocular nodes, such as \cite{waves,maxwell}. Motivated by their works, we propose to represent formation constraints using elliptic coordinates. Its coordinate lines are confocal ellipses and hyperbolae, as depicted in Fig. \ref{geometry}. The two coordinates, $\xi$ and $\eta$, indicate the distance and orientation of a point relative to the binocular system, separately. Given the pole locations $o_l=(-c,0)^T$ and $o_r=(c,0)^T$ in the \textit{xy}-plane, where $c\in {{\mathbb{R}}^{+}}$, the Cartesian coordinates are related to the elliptic coordinates by 
\begin{equation}
\begin{cases}
x = c \cosh(\xi)\cos(\eta) \\
y = c \sinh(\xi)\sin(\eta)
\end{cases}
\label{elliptic_2_xy}
\end{equation}
within the domain $D:=\{(\xi,\eta)^T|\xi\ge 0, 0\le \eta \le 2\pi\}$. To facilitate further discussions, we define $D^+,D^-\subset D$, where $D^+=\{(\xi,\eta)^T|\xi\ge 0, 0\le \eta \le \pi\}$, and $D^-=D-D^+$.

Without loss of generality, by assigning $o_l \leftarrow p_1 $ and $o_r \leftarrow p_2$, an elliptical coordinate system can be constructed by the two tracking vehicles in a binocular framework, where $c=d_{ij}/2$. Therefore, $p_t=(x,y)^T$ is alternatively represented by $^ep_t=(\xi,\eta)^T$, where the pre-superscript $e$ refers to elliptic coordinates. Then, we investigate how to obtain $^ep_t$ from $d_{it}$.

By eliminating $\xi$ of (\ref{elliptic_2_xy}) it derives
\begin{equation}
\sin^2(\eta)x^2 - \cos^2(\eta)y^2=c^2\sin^2(\eta)\cos^2(\eta),
\label{hyperbolae}
\end{equation}
which corresponds to a family of hyperbolae with the same focus distance $c$. If $\eta=k\pi/2$ for any $k\in \mathbb{Z}$, the hyperbola degenerates into a straight line. Then, by properties of hyperbola $\eta$ is calculated from $d_{it}$ as
\begin{equation}
\eta=
\begin{cases}
\arccos\frac{d_{1t}-d_{2t}}{2c}, \qquad \qquad if \quad ^ep_t \in D^+; \\
2\pi - \arccos\frac{d_{1t}-d_{2t}}{2c}, \qquad if \quad^ep_t \in D^-.
\end{cases}
\label{measurements_2_eta}
\end{equation}

Similarly, by eliminating $\eta$ of (\ref{elliptic_2_xy}) it derives
\begin{equation}
\sinh^2(\xi)x^2 + \cosh^2(\xi)y^2 = c^2\sinh^2(\xi)\cosh^2(\xi),
\label{ellipses}
\end{equation}
which corresponds to a family of ellipses with the same focus distance $c$. If $\xi=0$, the ellipse degenerates into a straight line. Then, by properties of hyperbola $\xi$ is calculated from $d_{it}$ as
\begin{equation}
\xi= {\rm arccosh}(\frac{d_{1t}+d_{2t}}{2c}).
\label{measurements_2_ksi}
\end{equation}

Together, (\ref{measurements_2_eta}) and (\ref{measurements_2_ksi}) constitute the transform equations from $d_{it}$ to $^ep_t$. To facilitate the stability analysis in the next section, we also present the transform equations from Cartesian coordinates $p_t$ to elliptic coordinates $^ep_t$ as follows.

By defining $p=\sin^2(\eta)$ and $q=-\sinh^2(\xi)$, we have $p\ge q$ since $0\le p \le 1$, $q\le 0$. According to \cite{Transform}, we have
\begin{equation}
p=\frac{-B+\sqrt{B^2+4c^2y^2}}{2c^2}, q=\frac{-B-\sqrt{B^2+4c^2y^2}}{2c^2},
\label{pq}
\end{equation}
where $B=x^2+y^2-c^2$.

As such, $\eta$ is recovered from the definition of $p$, depending on which quadrant the target is located, resulting in
\begin{equation}
\eta=
\begin{cases}
\eta_0, \qquad \qquad if \quad x\ge 0, y \ge 0; \\
\pi - \eta_0, \qquad if \quad x< 0, y \ge 0; \\
\pi + \eta_0, \qquad if \quad x\le 0, y<0; \\
2\pi - \eta_0, \qquad if \quad x> 0, y<0,
\end{cases}
\label{eta}
\end{equation}
where $\eta_0 = \arcsin \sqrt{p}$.

Similarly, $\xi$ is recovered from the definition of $q$ with a unique form as
\begin{equation}
\xi=\frac{1}{2}\ln(1-2q+2\sqrt{q^2-q}).
\label{xi}
\end{equation}

\subsection{Problem statement}

So far, we have established an elliptic coordinate system with the pole locations designated by the two tracking vehicles. In the following, we simply use $p_l$ and $p_r$ to represent the positions of the two vehicles instead of $p_1$ and $p_2$. Note that $^ep_t$ and $c$ completely characterize the shape and scale of the triangular formation in the collaborative tracking task. As such, we give a rigorous description of the binocular target tracking problem as follows.

\noindent \textbf {Problem 1 (Binocular target tracking).} Considering a pair of vehicles with agent model described by (\ref{eq_agentmodel}), for any given ${}^{e}p_{t}^{*}\in D$ and ${{c}^{*}}\in {{\mathbb{R}}^{+}}$, design a control law such that ${{}^{e}{p}_{t}}(t)\to {}^{e}p_{t}^{*}$ and $c(t)\to {{c}^{*}}$ as $t\to \infty $.

\section{Binocular Target Tracking}

\subsection{Tracking control law}

\begin{figure}
	\centering
	\includegraphics[width=2.6in]{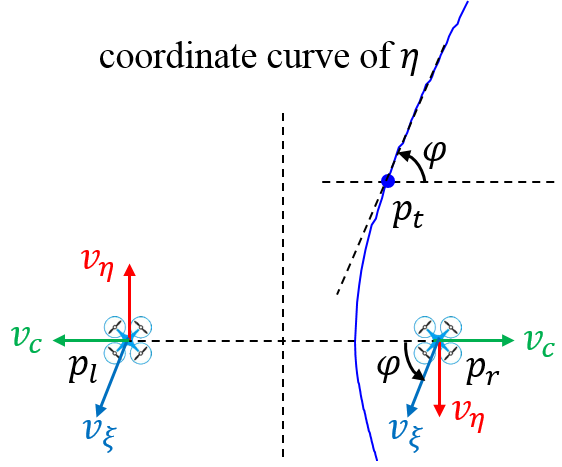}
	\caption{Combine collective velocities $v_\xi$, $v_\eta$ and $v_c$ to realize target tracking with formation constraints.}
	\label{vectors}
\end{figure}

The design of the tracking law obeys the \textit{binocular coordination principle}, which treats two vehicles as a joint system to control their collective behaviors. The control inputs are delivered to both tracking vehicles simultaneously. For a binocular system, it can be arbitrarily configured through three affine behaviors: translation, rotation, and scaling, as depicted in Fig. \ref{vectors}. To achieve target tracking with formation constraints, the translation speed $v_\xi$, rotation speed $v_\eta$, and scaling speed $v_c$ are designed to eliminate the tracking and formation errors in the directions/dimensions of $\xi$, $\eta$ and $c$, respectively. Therefore, the control component is given by
\begin{equation}
\begin{cases}
{{v}_{c}}={{\kappa }_{c}}(c^*-c) \\
{{v}_{\eta }}={{\kappa }_{\eta }}(\eta^*-\eta) \\
{{v}_{\xi}}={{\kappa }_{\xi}}(\xi^*-\xi)
\end{cases},
\label{eq_core}
\end{equation}
where ${{\kappa }_{c }}$, ${{\kappa }_\xi}$ and ${{\kappa }_\eta}$ are positive constants. 

Denote $R(\cdot)\in \mathbb{R}^{2 \times 2}$ as the rotation matrix of angle $(\cdot)$. The overall control inputs for the two vehicles are given by
\begin{equation}
\begin{cases}
u_l = A_l(p_r-p_l)  \\
u_r = A_r(p_r-p_l)
\end{cases},
\label{controller}
\end{equation}
where
\begin{equation}
\begin{cases}
A_l = \frac{1}{2c}\left(v_\xi R(\varphi) + v_\eta R(\pi/2)+v_c R(\pi) \right)  \\
A_r = \frac{1}{2c}\left(v_\xi R(\varphi) + v_\eta R(-\pi/2)+v_c R(0)\right)
\end{cases}.
\label{AlAr}
\end{equation}

\noindent \textbf {Remark 2.} Note that (\ref{eq_core}) is given in the form of a proportional controller. Such simplicity is due to the delicate definition of $\xi$ and $\eta$, where some new terms that encapsulate the coupling properties between the interacting vehicles in the Euclidean space, $B=x^2+y^2-c^2$ and $\sqrt{B^2+4c^2y^2}$, have emerged. As a result, the tracking errors are decoupled in the new space with clear physical meanings.

\subsection{Stability analysis for tracking a stationary target}
\label{stability_stationary}

Define the tracking error as
\begin{equation}
e=\left[ \begin{matrix}
2(c^2-(c^*)^2) \\
\eta - \eta^*  \\
\xi -\xi^*   \\
\end{matrix} \right]
=\left[ \begin{matrix}
e_1 \\
e_2 \\
e_3
\end{matrix} \right],
\label{error}
\end{equation}
where $c=(p_l-p_r)^T(p_l-p_r)/2$. The error dynamics is derived as
\begin{equation}
\dot e=\left[ \begin{matrix}
(p_l-p_r)^T(\dot{p_l}-\dot{p_r})  \\
\dot \eta  \\
\dot \xi   \\
\end{matrix} \right]
=\left[ \begin{matrix}
\dot {e_1} \\
\dot {e_2} \\
\dot {e_3}
\end{matrix} \right].
\label{errordynamics}
\end{equation}

Now we are ready to present the first main result.

\noindent \textbf {Theorem 1.} Consider a pair of tracking vehicles described by (\ref{eq_agentmodel}). Then, for any given $^ep_{t}^{*}\in D$ and ${{c}^{*}}\in {{\mathbb{R}}^{+}}$, by applying the proposed controller (\ref{eq_core}) to (\ref{AlAr}), $e=0_3$ is an exponentially stable equilibrium point of the tracking error dynamics (\ref{errordynamics}). 

\noindent \textbf {Proof.}
Consider the Lyapunov function candidate
\begin{equation}
V = \frac{1}{2}e^Te = \frac{1}{2}(e_1^2+e_2^2+e_3^2).
\label{Lyapunov}
\end{equation}

The time-derivative of $V$ is
\begin{equation}
\dot V = \dot e^Te = 
e_1\dot e_1+
e_2\dot e_2+
e_3\dot e_3.
\label{dotLyapunov}
\end{equation}

Next, we will discuss the dynamics of $e_1$, $e_2$ and $e_3$, separately, to show the convergence of the tracking errors.

$\bullet$ \textit {Step 1:} analysis on $\dot e_1$

Substituting the obtained expressions in (\ref{errordynamics}) with the proposed controller (\ref{eq_core}) to (\ref{AlAr}) yields a quadratic form
\begin{equation}
\dot e_1 = (p_l - p_r)^T(A_r - A_l)(p_l - p_r),
\label{dote1}
\end{equation}
where
\begin{equation}
A_r - A_l=\frac{1}{c}(v_cI_2+\left[ \begin{matrix} 
0 & -1  \\
1 & 0  \\
\end{matrix} \right]).
\label{Ar_minus_Al}
\end{equation}

Note that the second item in (\ref{Ar_minus_Al}) is a skew symmetric matrix, whose quadratic form is 0. Utilizing $c=\left\| {{p}_{l}}-{{p}_{r}} \right\|/2$ and $v_c = \kappa_c (c^*-c)$, (\ref{dote1}) is further simplified to
\begin{equation}
\dot e_1 = 4c\kappa_c(c^*-c)=-\frac{2c\kappa_c}{c+c^*}e_1:=-F_1e_1,
\label{reduced_dote1}
\end{equation}
where $F_1$ is positive definite for $c>0$, so that $e_1(t) \to 0$ as $t \to \infty$, which infers $c(t) \to c^*$ monotonously. Therefore, $c(t)\ge\min\{c(t_0),c^*\}>0$ for all $t\ge t_0$.
Moreover, $F_1$ is a function of $e_1$ according to (\ref{errordynamics}), and the error dynamics given by (\ref{reduced_dote1}) constitute an autonomous system. Note that $F_1$ is a class $\mathcal{K}$ function with repect to $e_1$. Thus, we have $F_1>k_1$, where
\begin{equation}
k_1 = 2\min\left \{\frac{2\kappa_c \sqrt{\frac{e_1(t_0)}{2}+(c^*)^2}}{\sqrt{\frac{e_1(t_0)}{2}+(c^*)^2}+c^*},\kappa_c\right \}
\label{k1}
\end{equation}
is a positive constant.

$\bullet$ \textit {Step 2:} analysis on $\dot e_2$

By partial differentiation of $\eta$ and combining with (\ref{errordynamics}), the following expression is obtained:
\begin{equation}
\begin{split}
\dot e_2 &= \frac{\partial \eta}{\partial x}\dot x+\frac{\partial \eta}{\partial y}\dot y \\
&= {\rm sgn}(xy)G_2\left((\frac{B}{G_1}-1)x\dot x + (\frac{B+2c^2}{G_1}-1)y\dot y\right),
\end{split}
\label{dot_e2}
\end{equation}
where $G_1=\sqrt{B^2+4c^2y^2}$, $G_2=\frac{1}{2c^2\sqrt{p-p^2}}$. The sign function ${\rm sgn}(xy)$ is used because $\eta$ has different definitions in different quadrants.

The relative target motion dynamics by the proposed controller is given by
\begin{equation}
\left[ \begin{matrix} 
\dot x  \\
\dot y  \\
\end{matrix} \right]={\rm sgn}(xy)\frac{v_\xi}{G_\varphi}
\left[ \begin{matrix} 
y\cos^2\eta  \\
x\sin^2\eta  \\
\end{matrix} \right]+
\frac{v_\eta}{c}\left[ \begin{matrix} 
-y  \\
x \\
\end{matrix} \right],
\label{dotxy_formation}
\end{equation}
where $G_\varphi=\sqrt{x^2\sin^4\eta+y^2\cos^4\eta}>0$. The sign function $\rm{sgn}(xy)$ is used to indicate the direction of vehicle movements when the target is in different quadrants.

Substituting (\ref{dotxy_formation}) in (\ref{dot_e2}), combining with $\sin^2\eta=(G_1-B)/(2c^2)$ and $v_\eta=-\kappa_\eta e_2$ yields
\begin{equation}
\dot e_2=-2c\kappa_\eta\vert xy \vert \frac{G_2}{G_1}e_2:=-F_2e_2,
\label{dot_e2_3}
\end{equation}

We prove that there exists a positive constant $0<\mu<c$ such that $F_2 \ge k_2$, where $k_2 = \frac{\kappa_\eta}{c}\sqrt{\frac{\mu^2}{\mu^2+c^2}}$ is a positive constant (see proof A in Appendix).

$\bullet$ \textit {Step 3:} analysis on $\dot e_3$

Similarly, by partial differentiation of $\xi$ and combining with (\ref{errordynamics}), we have 
\begin{equation}
\begin{split}
\dot e_3 &= \frac{\partial \xi}{\partial x}\dot x+\frac{\partial \xi}{\partial y}\dot y \\
&= G_3\left((1+\frac{B}{G_1})x\dot x + (1+\frac{B+2c^2}{G_1})y\dot y\right),
\end{split}
\label{dot_e3}
\end{equation}
where $G_3=\frac{1}{2c^2\sqrt{q^2-q}}$.

Substituting (\ref{dotxy_formation}) again and combining with $v_\eta=-\kappa_\eta e_2$, $v_\xi=-\kappa_\xi e_3$, we have
\begin{equation}
\dot e_3 = -2\vert xy \vert \kappa_\xi \frac{G_3}{G_\varphi}e_3-2cxy\kappa_\eta\frac{G_3}{G_1}e_2.
\label{dot_e3_3}
\end{equation}

As discussed above, the second term is exponentially convergent. Therefore, we only investigate the convergence of the first term. Define $F_3=2\vert xy \vert \kappa_\xi\frac{G_3}{G_\varphi}$. We prove that there exists a positive $\nu>c$ such that $F_3\ge k_3$, where $k_3 = \frac{\kappa_\xi}{c^2\sqrt{\nu^2+c^2}}$ is a positive constant (see proof B in Appendix).

$\bullet$ \textit {Step 4:} Combining Steps 1-3, we have $\dot V < -k_4 V$, where $k_4:=2\min\{k1,k2,k3\}$ is a positive constant. Then, by (\cite{Khalil}, Theorem 4.10), one concludes that the origin $e=0$ is exponentially stable, and the proof is complete. \qed

\noindent \textbf {Remark 3.} In the above discussions, we exclude the case where $-c\le x \le c, y=0$. However, it is easy to check that such a case would not result in singularity. For example, if $y=0$, $y^*> 0$, then it is equivalent to $\xi=0$, $\xi^*>0$, leading to a positive $v_\xi$, which will drive the vehicles away from \textit{x}-axis so that $y\neq 0$ immediately. 

\subsection{Tracking a moving target under slow drift}
So far we have established the exponential stability for the tracking system for a stationary target, where $\dot p_t^* = 0$. Now we consider the tracking control with a moving target $p_t^*(t)$, by assuming that there exists $\varepsilon \in [0,\infty)$ such that for all $t \in \mathbb{R}$,
\begin{equation}
\left\| \dot p_t^*(t) \right\| \le \varepsilon.
\label{dot_target}
\end{equation}

Herein, we regard the drift-free error system (\ref{errordynamics}) as the nominal system and rewrite it as $\dot e = f(t,e)$. The Lyapunov function of the norminal system $V(t,e):[0,\infty)\times D \to \mathbb{R}^+$ is defined by (\ref{Lyapunov}), where $D=\{e\in \mathbb{R}^3|\left\| e \right\|<r\}$. Then, the moving target's slow drift is equivalently regarded as an additional translational movement of the binocular system due to the motion relativity, which acts as a drift term $g(t,e)$ additive to the nominal system, resulting in a perturbed tracking error system
\begin{equation}
\dot e = f(t,e) + g(t,e).
\label{pertubed_system}
\end{equation}

The following result is herein presented.

\begin{figure*}
	\centering
	\includegraphics[width=6.8in]{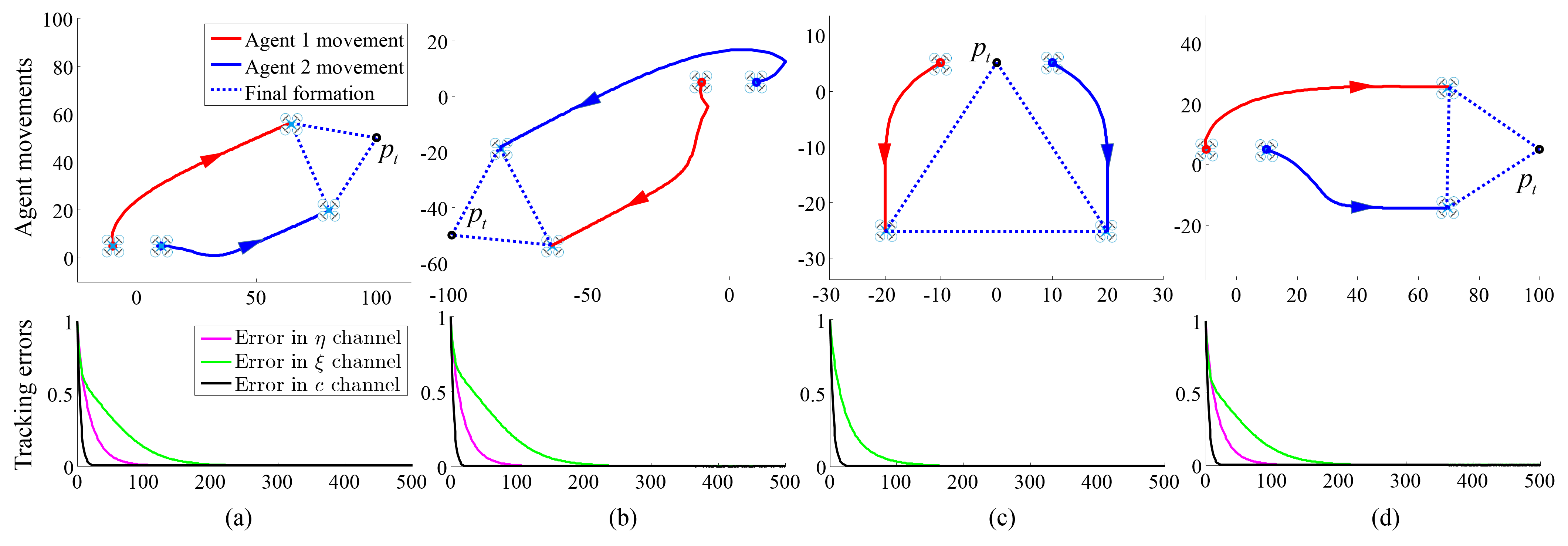}
	\caption{Coordinated tracking of a stationary target under different initial positions: (a) $p_t\in D^+$; (b) $p_t\in D^-$; (c) $p_t$ is located on the midpoint of two vehicles; (d) $p_t$ is on the x-axis and $x>c$. The upper figures show the tracking vehicles' trajectories and the final formations in Euclidean space. Figures in the below show the tracking errors of individual channels, which all converge to 0 exponentially fast.}
	\label{fig_result_1}
\end{figure*}

\noindent \textbf {Theorem 2.} Consider the perturbed system described by (\ref{pertubed_system}) with a slow-moving target constainted by (\ref{dot_target}). Then, for all $\left\| e(0) \right\|<r$, there exists a positive constant $K$ such that 
\begin{equation}
\underset{t\to \infty }{\mathop{\lim }} \left\| e(t) \right\|\le K\varepsilon.
\nonumber
\end{equation}

\noindent \textbf {Proof.} The target dynamics in the local $xy$-frame with respect to a stationary binoculr system is described by
\begin{equation}
\begin{cases}
\dot x_t = v \cos \theta_t  \\
\dot y_t = v \sin \theta_t
\end{cases},
\label{target_dynamics}
\end{equation}
where $v=\left\| \dot p_t^*(t) \right\|$,$\theta_t\in[0,2\pi] $ denotes the direction of $\dot p_t^*(t)$.

Specifically, the drift term can be equivalently described by a vector given by $g(t,e)=[0,g_\eta(t,e), g_\xi(t,e)]^T$. Then, by following a procedure similar to Subsection \ref{stability_stationary}, we will derive the expressions and upper bounds of $ g_\eta(t,e)$ and $ g_\xi(t,e)$ separately to investigate the upper bound of $\vert g(t,e) \vert$.

Substituting (\ref{target_dynamics}) in (\ref{dot_e2}) yields
\begin{equation}
g_\eta(t,e) = G_2\left((\frac{B}{G_1}-1)x\dot x_t + (\frac{B+2c^2}{G_1}-1)y\dot y_t\right).
\label{g_eta}
\end{equation}

Applying Cauchy's inequality and combining with (\ref{dot_target}), we have
\begin{equation}
\begin{split}
\vert g_\eta(t,e) \vert &\le vG_2\sqrt{(\frac{B}{G_1}-1)^2x^2+(\frac{B+2c^2}{G_1}-1)^2y^2} \\
&\le \frac{\varepsilon}{c\sqrt{h}},
\end{split}
\label{g_eta_norm}
\end{equation}
where $h=G_1/c^2$.
Similarly, substituting (\ref{target_dynamics}) in (\ref{dot_e3}) yields
\begin{equation}
\begin{split}
g_\xi(t,e) &= G_3\left((1+\frac{B}{G_1})x\dot x_t + (1+\frac{B+2c^2}{G_1})y\dot y_t\right).
\end{split}
\label{g_ksi}
\end{equation}

By applying Cauchy's inequality again and combining with (\ref{dot_target}), we have
\begin{equation}
\begin{split}
\vert g_\xi(t,e) \vert &\le vG_3\sqrt{(1+\frac{B}{G_1})^2x^2+(1+\frac{B+2c^2}{G_1})^2y^2} \\
&\le \frac{\varepsilon}{c\sqrt{h}}.
\end{split}
\label{g_ksi_norm}
\end{equation}

Note that (\ref{g_eta_norm}) and (\ref{g_ksi_norm}) share the same expression, which has the same monotonicity property with $F_3$. Then, suppose that $(x-c)^2+y^2\le\delta^2$, where $0<\delta<c$, we have $\vert g_\eta(t,e) \vert\le \bar K \varepsilon$ and $\vert g_\xi(t,e) \vert\le \bar K \varepsilon$, where $\bar K=1/\sqrt{\delta^2-2c\delta}$.

Now we have proven that there exists $0<\theta<1$ and $\bar K\in \mathbb{R}^+$ such that 
\begin{equation}
\left\| g(t,e) \right\|\le \bar K \varepsilon < k_4\theta r
\label{g_norm}
\end{equation}

Then, the result immediately follows by applying (\cite{Khalil}, Lemma 9.2), which describes the bounded stability of perturbed systems, and we have $K=\bar K/(k_4\theta)$. \qed

\noindent \textbf {Remark 4.} Theorem 2 shows that the proposed controller is able to track a moving target and the tracking errors converge to a small region proportional to the target speed, which indicates that small perturbation will not result in large steady-state deviations from the origin. In fact, we only require the target speed to not exceed a sphere without explicit restrictions on the moving trajectory (\ref{dot_target}), and a specific moving target can be regarded as a special case of (\ref{dot_target}).

\begin{figure}
	\centering
	\includegraphics[width=3.3in]{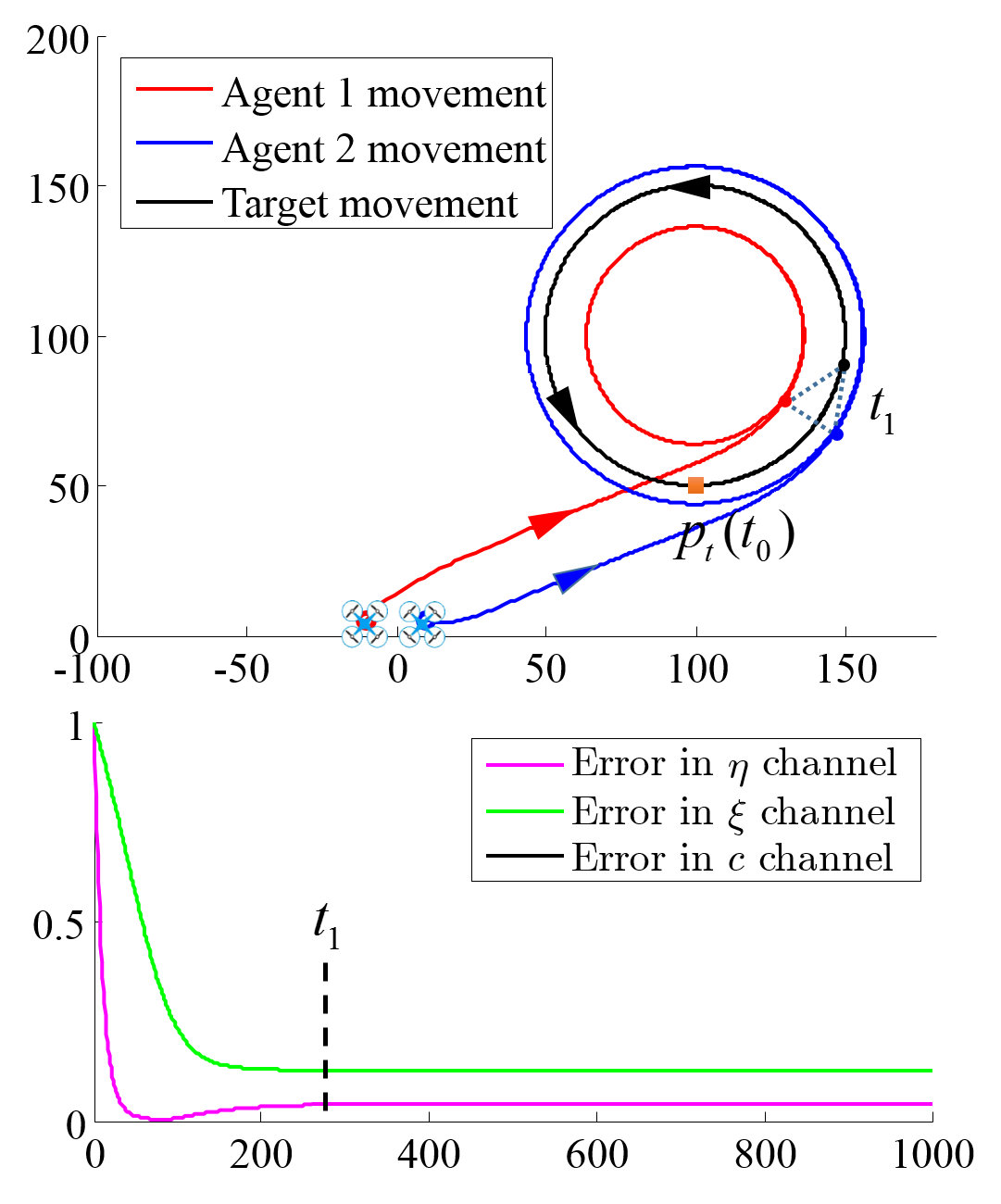}
	\caption{Coordinated tracking of a circular moving target with $^ep_t^*=(1.2,\pi/2)^T$ and $c^*=20$. After $t_1$, the tracking error converges to a small interval proportional to the target speed.}
	\label{fig_result_2}
\end{figure}

\section{Results and Discussions}

In this section, we show several examples to illustrate the tracking performance of the proposed controller.

\subsection{Tracking control with a stationary target}
The first example illustrates how the tracking vehicles moving from an initial position to collaboratively track an arbitrary placed stationary target with formation constraints. In all simulations, the initial positions are set as ${{p}_{1}}(t_0)={{(-10,5)}^{T}}$ and ${{p}_{2}}(t_0)={{(10,5)}^{T}}$. The desired formation configuration is given by $^ep_t^* = (1.2,\pi/2)^T$ and $c^*=40$. The parameters are set as $\kappa_\xi=1.0$, $\kappa_\eta=1.0$ and $k_c = 0.1$. 

Fig. \ref{fig_result_1} shows four rounds of simulations with different target positions $p_t$. As expected, in all the simulations the system converges to the desired formation w. r. t. the target, and the tracking errors converge to 0 exponentially fast. This demonstrates the global convergence of the tracking system with the proposed controller. The tracking errors provide an intuitive description of the underlying mechanism.

\subsection{Tracking control with a moving target}

Now we consider tracking a moving target with an initial position $p_t(t_0)=(100,50)^T$. In the first example, the target circles around the point $(100,100)^T$ with a radius of 50 and a speed of 5. The values of the setup $p_1(t_0)$, $p_2(t_0)$, $^ep_t^*$ and $c^*$ are the same as those in the above subsection. The results are illustrated in Fig. \ref{fig_result_2}. Note that the tracking vehicles and the target eventually approach the desired formation after $t_1$. As the time approaches infinity, the tracking errors converge to a small interval, as stated in Theorem 2.

\begin{figure}
	\centering
	\includegraphics[width=3.3in]{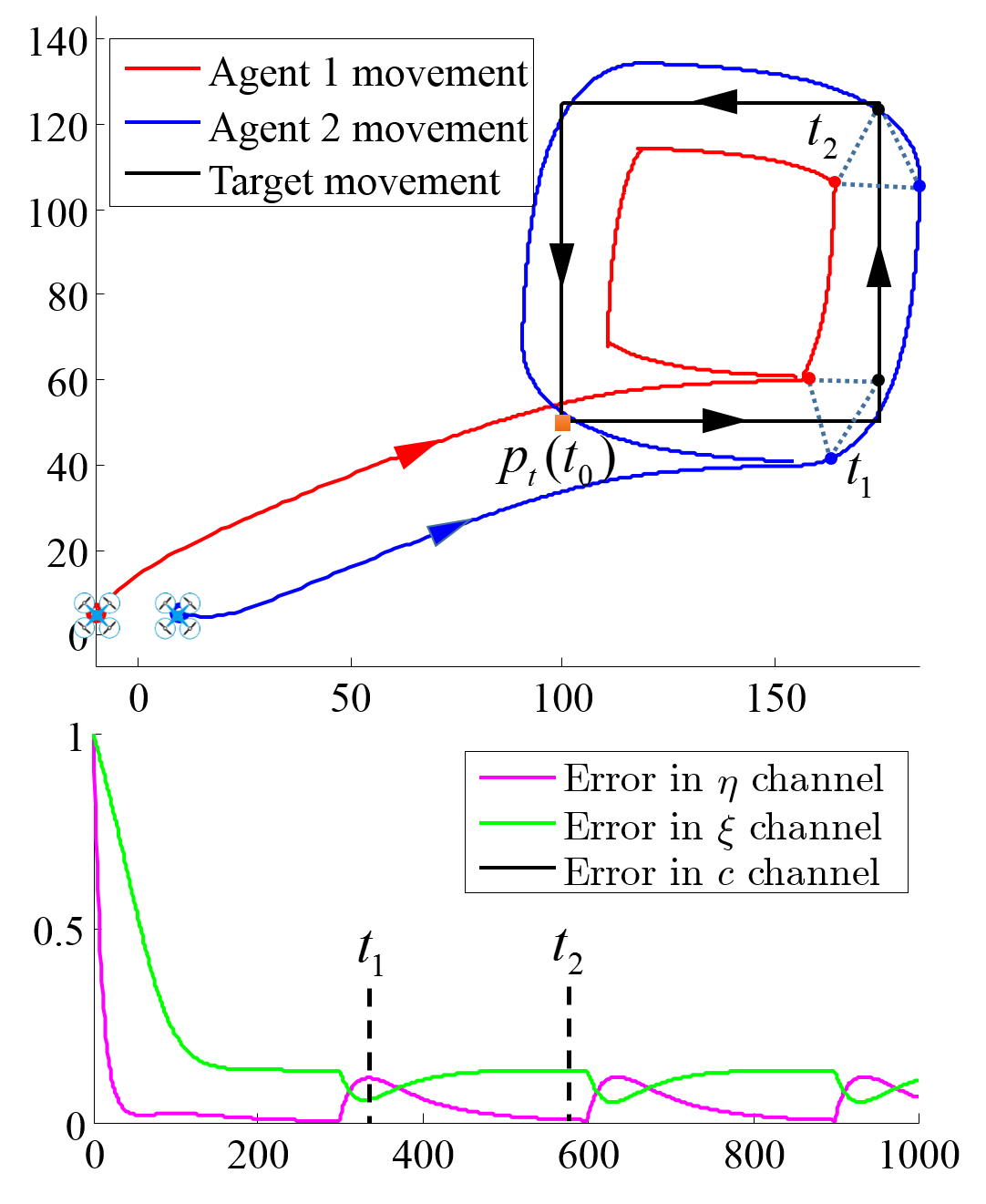}
	\caption{Coordinated tracking of a moving target trajectory with sharp turns under $^ep_t^*=(1.2,\pi/2)^T$ and $c^*=20$. The target trajectory is continuous in stages, resulting in some oscillations in the tracking errors but eventually converge to a limited interval.}
	\label{fig_result_3}
\end{figure}

In the next example, we consider a target trajectory with sharp turns, which is common in adversarial tracking tasks. The results are illustrated in Fig. \ref{fig_result_3}. Note that the tracking vehicles actively adjust their movements to track the target. The tracking errors converge to a small value, which can be further reduced by adjusting control gains. 

These two examples demonstrate that the proposed controller enables successful tracking of a moving target with formation constraints.

\section{Concluding Remarks}

In this paper, we have proposed a novel control method for collaborative target tracking using a pair of tracking vehicles. The formation constraints in the tracking task are represented in elliptic coordinates and the control objective is achieved by regulating the rotation, translation, and scaling of the target tracking system. A detailed stability analysis, as well as a rich set of simulations, have been provided to demonstrate the tracking performance with a stationary or slow-drifting target. The proposed collaborative target tracking control approach can also be extended to a multi-robot system, which will be addressed in the future.

\section{Appendix}

\noindent \textbf {Proof A} (lower bound of $F_2$). Recall that $F_2 = 2c\kappa_\eta\vert xy \vert \frac{G_2}{G_1}$. Let $x^2+y^2=gc^2$, $G_1=hc^2$, it is easy to check that
$g-1<h<g+1$. Substituting $B$ and $G_1$ with $h$ and $g$ yields
\begin{equation}
\begin{split}
F_2&=2c\kappa_\eta \frac{\vert xy \vert}{\sqrt{(G_1-B)(2c^2+B-G_1)}G_1} \\
&=\frac{c\kappa_\eta}{2}\frac{\sqrt{(g+1)^2-h^2}\sqrt{h^2-(g-1)^2}}{\sqrt{(h+1-g)(g+1-h)}hc^2} \\
&=\frac{\kappa_\eta}{2c}\sqrt{1+(\frac{g}{h})^2+2\frac{g}{h}-\frac{1}{h^2}}.
\end{split}
\nonumber
\end{equation}

Since $F_2$ is symmetrical along the y-axis and the x-axis, it is sufficient to consider its monotonicity in the first quadrant.

If $x^2+y^2\ge c^2$, $g \ge 1$, then $F_2 \ge \frac{\kappa_\eta}{2c}\sqrt{1+\frac{2}{h}} \ge \frac{\kappa_\eta}{2c}$. Besides, note that when $g\to \infty$, it follows that $g/h \to 1$ and $1/h^2 \to 0$, thus $F_2 \to \kappa_\eta/c$.

If $x^2+y^2<c^2$, by partial differentiation of $F_2$ with respect to $x$, it derives $\frac{\partial F_2}{\partial x}=M_2xN_2$, where $M_2=\frac{\kappa_\eta}{c^3h\sqrt{(g+h)^2-1}}$ and $N_2 = (h^2-(g-1)^2)(h+g)+(1-g)(g-1+h)$. It is easy to check that both $M_2$ and $N_2$ are positive, thus $\frac{\partial F_2}{\partial x}>0$. Then, by partial differentiation of $F_2$ with respect to $y$, it follows $\frac{\partial F_2}{\partial x}=M_2y(N_2+2c^2(1-g^2-gh))$. Since $h<g+1$, the last term $2c^2(1-g^2-gh)>2c^2(1-g)>0$. Therefore, $\frac{\partial F_2}{\partial y}>\frac{\partial F_2}{\partial x}>0$. As such, $F_2$ is monotonic in both directions. Suppose there exists positive constants $0<\mu<c$ that satisfies $y\ge \mu$, we have $F_2 \ge k_2$, where $k_2 = \frac{\kappa_\eta}{c}\sqrt{\frac{\mu^2}{\mu^2+c^2}}$ is a positive constant.

\noindent \textbf {Proof B} (lower bound of $F_3$). Recall that $F_3=2\vert xy \vert \kappa_\xi\frac{G_3}{G_\varphi}$. By substituting $B=(g-1)c^2$ and $G_1=hc^2$, it is simplified as $F_3=\frac{\kappa_\xi}{c^3\sqrt{h}}$. Since $F_3$ is symmetrical along the y-axis and the x-axis, it is sufficient to investigate its monotonicity in the first quadrant. It is obvious to derive that $\frac{\partial F_3}{\partial x} = -\frac{\kappa_\xi xB}{c^5h\sqrt{h}G_1}$. If $x^2+y^2<c^2$, $\frac{\partial F_2}{\partial x}>0$. If $x^2+y^2>c^2$, $\frac{\partial F_2}{\partial x}<0$. As to the \textit{y}-direction, $\frac{\partial F_3}{\partial y} = -\frac{\kappa_\xi y(B+2c^2)}{c^5h\sqrt{h}G_1}<0$. Therefore, for a bounded target position $x^2+y^2\le \nu$, where $\nu>c$, we have $F_3\ge k_3$, where $k_3 = \frac{\kappa_\xi}{c^2\sqrt{\nu^2+c^2}}$ is a positive constant.

\addtolength{\textheight}{-7cm}   % This command serves to balance the column lengths

%%%%%%%%%%%%%%%%%%%%%%%%%%%%%%%%%%%%%%%%%%%%%%%%%%%%%%%%%%%%%%%%%%%%%%%%%%%%%%%%

%%%%%%%%%%%%%%%%%%%%%%%%%%%%%%%%%%%%%%%%%%%%%%%%%%%%%%%%%%%%%%%%%%%%%%%%%%%%%%%%

%%%%%%%%%%%%%%%%%%%%%%%%%%%%%%%%%%%%%%%%%%%%%%%%%%%%%%%%%%%%%%%%%%%%%%%%%%%%%%%%

%%%%%%%%%%%%%%%%%%%%%%%%%%%%%%%%%%%%%%%%%%%%%%%%%%%%%%%%%%%%%%%%%%%%%%%%%%%%%%%%

\end{document}